\documentclass[aps, prl,nofootinbib,tightenlines, twocolumn,floatfix]{revtex4}

 \usepackage[dvips,final]{graphicx}
  \usepackage{amssymb}
   \usepackage{amsmath}
    \usepackage{amsfonts}
     \usepackage{epsfig}
      \usepackage{bm} 
\makeatletter\AtBeginDocument{\let\@elt\relax}\makeatother

\usepackage{mathrsfs}
\usepackage{slashed}

\usepackage[colorlinks = true,
linkcolor = blue,
urlcolor  = blue,
citecolor = blue,
anchorcolor = blue]{hyperref}

\begin{document}

\title{Causality, the Kovtun-Son-Starinets bound, and a novel sum rule for spectral densities}

\author{G. Yu. Prokhorov$^{\, a,b}$}
\email{prokhorov@theor.jinr.ru}

\author{O. V. Teryaev$^{\, a,b}$}
\email{teryaev@jinr.ru}


\affiliation{$^{a\,}$Joint Institute for Nuclear Research, Joliot-Curie 6, Dubna, 141980, Russia}
\affiliation{$^{b\,}$NRC Kurchatov Institute, Moscow, Russia}

\begin{abstract}
We directly show that the local ratio of the shear viscosity to the entropy density for Unruh radiation at a finite distance from the horizon is universal and satisfies the relation $ \eta/s = 1/(4\pi c_s^2) $, which involves the speed of sound $ c_s $. Since $ c_s^2 \leq 1 $ by causality, this establishes the close connection between the famous Kovtun-Son-Starinets bound and causality. Moreover, we show that the ratio of bulk to shear viscosity saturates another well-known bound for the bulk viscosity, predicted within holographic approach. We also show that the condition of isotropy of thermal radiation in the Rindler space leads to a novel sum rule relating the $ c^{(0)}(\mu) $ and $ c^{(2)}(\mu) $ spectral densities, and we explicitly demonstrate its validity for conformal field theory and free massive Dirac fields in any number of dimensions. The sum rule provides the validity of Pascal law and bears some similarity with Burkhardt-Cottingham sum rule for spin-dependent parton distributions. Our result suggests a new perspective on dissipative transport phenomena in media undergoing extreme acceleration, such as quark–gluon plasma created in relativistic heavy-ion collisions.
\end{abstract}

\maketitle

\section{Introduction}
\label{sec_intro}

A deep connection between the properties of matter and gravity has become one of the central paradigms of fundamental physics. A striking example of this relationship, established using the holographic principle, is the celebrated Kovtun-Son-Starinets (KSS) bound on the ratio of shear viscosity to entropy density \cite{Policastro:2001yc, Kovtun:2004de, Chiofalo:2025jph}
\begin{eqnarray}
\frac{\eta}{s}\geq \frac{\hbar}{4\pi k_B}\,.
\label{kss}
\end{eqnarray}
In this paper, we establish -- outside the holographic context -- a direct connection between the KSS bound and causality. Namely, by analyzing thermal radiation from a causal horizon, we show that (\ref{kss}) holds as long as the speed of sound remains below the speed of light. Interestingly, this extends a similar link between causality and viscosity, found in holographic models with Gauss-Bonnet gravity dual \cite{Brigante:2008gz}, where a weaker viscosity bound is broken by superluminal gravitons.

Historically, the saturation of the KSS (prior to its proof in holographic framework) bound was first demonstrated just for black holes within the membrane paradigm \cite{Damour:1979wya, Parikh:1997ma}. Some time ago, in \cite{Chirco:2010xx} followed  and generalized recently in \cite{Lapygin:2025zhn}, the saturation  was demonstrated by direct application of linear-response theory to the thermal radiation in Rindler space -- the simplest flat-space analog of a black hole \cite{Unruh:1976db, Bisognano:1976za, Witten:2024upt}. As a result, it appeared, surprisingly, that the radiation itself has some finite viscosity saturating the bound. This may imply an intriguing correspondence between the properties of a classical membrane and the quantum radiation above it. It may be also interesting to compare this with the seminal results for entanglement entropy \cite{Srednicki:1993im, Bombelli:1986rw}. Given the origin of the thermal radiation, this viscosity was called \cite{Chirco:2010xx} the ``entanglement viscosity''.

In \cite{prep1}, the results of \cite{Chirco:2010xx, Lapygin:2025zhn} were further generalized to a broad class of field theories, providing explicit formulas for the entanglement viscosity in terms of the spectral functions of the universal spectral representation. This provided an additional support of the deep relationship \cite{Zamolodchikov:1986gt, Cappelli:1990yc} between irreversibility and unitarity.

In this paper, we address the local transport properties at a given point above the horizon, and show that the ratio of entanglement viscosity to entropy density satisfies the relation\footnote{It is interesting to notice that if we consider integrated (and therefore two-dimensional) quantities with an appropriate regularization near the horizon the entanglement viscosity "globally" saturates the KSS bound \cite{Chirco:2010xx, Lapygin:2025zhn, prep1}. This demonstrates the universality of (\ref{kss}), which is realized in different ways for "local" (three-deimensional) and global (two-dimesnsional) quantities. }
\begin{eqnarray}
\frac{\eta}{s}\Big|_{loc}= \frac{\hbar c^2}{4\pi k_B c_s^2}\,,
\label{main}
\end{eqnarray}
which includes the ratio $c/c_s$ of speed of light to the speed of sound \footnote{Recall, that KSS bound does not depend on the speed of light}. We show the validity of (\ref{main}) for a wide class of theories, including non‑conformal and interacting ones, in any number of dimensions, and briefly discuss possible phenomenological implications. It is instructive to note that, in our framework, (\ref{main}) is equivalent to another relation
\begin{eqnarray}
\frac{\eta}{c_V}\Big|_{loc}= \frac{\hbar}{4\pi k_B}\,,
\label{main1}
\end{eqnarray}
where $ c_V $ is the specific heat at constant volume. 

Comparing (\ref{main}) and (\ref{kss}), one can immediately see that for the KSS bound to be violated, the speed of sound should become greater than the speed of light $ c_s > c $, which would mean a violation of causality \cite{Ellis:2007ic, Kovtun:2019hdm}. Thus, (\ref{main}) indicates the relationship between the bound (\ref{kss}) and the causality.

As another important consequence, we find that the entanglement bulk viscosity $ \zeta $ saturates the well-known bound also proposed in holographic approaches \cite{Buchel:2007mf}
\begin{eqnarray}
\frac{\zeta}{\eta}\Big|_{loc} = 2 \left(\frac{1}{d-1}-\frac{c_s^2}{c^2}\right) \,.
\label{zeta_eta_intro}
\end{eqnarray}

These derivations rely on an observation regarding the isotropy of the thermal radiation. In general, Unruh radiation can be anisotropic with different longitudinal and transverse pressures. However, by imposing the isotropy condition, we derive a novel sum rule relating the spectral functions $c^{(0)}(\mu)$ and $c^{(2)}(\mu)$ from the correlator of two stress-energy tensors, and explicitly verify its validity for any conformal field theory, as well as for free massive Dirac fields in any number of dimensions.

We use the system of units $ e=\hbar=c=k_B=1 $, and the metric signature $ (+,-,-,-) $, restoring the dimensionful quantities where it is appropriate.

\section{Isotropy sum rule for spectral functions}
\label{sec_sum}
\subsection{Thermodynamics in Rindler space}
\label{secsub_therm}

A medium with a proper temperature $ T $ and acceleration $ a $ can be described by considering quantized fields in the Euclidean Rindler space. We will consider the general $ d $-dimensional case
\begin{eqnarray}
ds^2 = \rho^2 d\tau^2+(dx^1)^2+...+(dx^{d-2})^2+d\rho^2\,,
\label{rindler}
\end{eqnarray}
where the coordinates have a hydrodynamic interpretation: the period of the Euclidean time $ \tau $ is determined by the inverse proper temperature $ 1/T $, and the radial coordinate defines the inverse acceleration $ \rho=1/a $.

This manifold $\mathcal{M} = \mathbb{R}^{d-2} \otimes\, \mathcal{C}^2_{\nu}  $ contains a 2D cone $  \mathcal{C}^2_{\nu} $ and is an example of a well-known conical space \cite{Dowker:1994fi, Solodukhin:2011gn}. The angular deficit  $ 2\pi (1-\nu^{-1}) $ of the cone is described by the parameter
\begin{eqnarray}
\nu= 2 \pi T/a \,.
\label{nu}
\end{eqnarray}
Considering quantum fields living in the space (\ref{rindler}), one can find the mean value of the energy-momentum tensor as a function of acceleration and temperature\footnote{After calculating the averages, it is necessary to restore  the Lorentz metric signature.}
\begin{eqnarray}
\langle \hat{T}^{\mu}_{\nu} \rangle (a,T)=\text{diag} (\varepsilon, -p_\perp,...,-p_{||}) \,.
\label{Tgen}
\end{eqnarray}
In (\ref{Tgen}) symmetry guarantees EMT to be diagonal and having the equal transverse components of pressure \cite{Buzzegoli:2017cqy}. When the temperature is equal to the Unruh temperature the cone in (\ref{rindler}) unfolds into a plane and (\ref{Tgen}) vanishes due to the normalization to the Minkowski vacuum
\begin{eqnarray}
\langle \hat{T}^{\mu}_{\nu} \rangle (a,T=T_U)=0\,,\quad T_U=\frac{a}{2\pi}\,,
\label{unruh}
\end{eqnarray}
which is a manifestation of the well-known Unruh effect, that is, the Minkowski vacuum is perceived by an accelerated observer as a thermal bath with a temperature $ T_U $ \cite{Becattini:2017ljh, Unruh:1976db, Bisognano:1976za}. Let us emphasize that, in general, acceleration (or $ T_U $) and temperature $ T $ are two independent parameters that determine the concrete state of the system. To illustrate (\ref{Tgen}) and (\ref{unruh}), we present a well-known result for massless Dirac fields in four dimensions \cite{Frolov:1987dz, Dowker:1994fi, Prokhorov:2019cik}
\begin{eqnarray} \nonumber
\langle \hat{T}^{\mu}_{\nu} \rangle (a,T)&=& \left(\frac{7\pi^2 T^4}{60}+\frac{T^2 a^2}{24}-\frac{17 a^4}{960 \pi^2}\right) \\ &&\cdot \,\text{diag}(1,-1/3,-1/3,-1/3)\,.
\label{12}
\end{eqnarray}
In particular, it is easy to see that (\ref{12}) vanishes at $ T=T_U $.

The first derivatives with respect to $ T $ determine the specific heat, entropy density and speed of sound\footnote{There are various approaches to find the entanglement entropy \cite{Solodukhin:2011gn}, and we use the one, based on the relativistic spin hydrodynamics \cite{Becattini:2023ouz, Lapygin:2025zhn, prep1}. Also in the considered case with zero chemical potentials $ \mu=0 $, the squared speed of sound can be defined as $c_s^2 = \frac{\partial p/\partial T}{\partial \varepsilon/\partial T}$ \cite{Laine:2016hma, Cleymans:2011fx}, where we fix the acceleration $a_{\nu}$ as constant during sound wave propagation.}
\begin{eqnarray} \nonumber
c_V(a,T)&=&\frac{\partial \varepsilon (a,T)}{\partial T} \Big|_{a}\,, \\ \nonumber
s(a,T) &=&\frac{\partial p (a,T)}{\partial T} \Big|_{a} \,, \\
c_s^2(a,T) &=& \frac{\partial p}{\partial \varepsilon} \Big|_{a} =  \frac{\partial p(a,T)/\partial T}{\partial \varepsilon(a,T)/\partial T}=\frac{s}{c_V}\,.
\label{defder}
\end{eqnarray}
It is not difficult, for example, to find $ c_V, s $ and $ c_s^2 $ for massless Dirac field from (\ref{defder}), (\ref{Tgen}) and (\ref{12}).


Below we will consider the case of general quantum field theory (including non-conformal and interacting ones) in an arbitrary number of dimensions.

\subsection{Isotropy sum rule}
\label{secsub_sum}

When $ \nu=1 $, the space (\ref{rindler}) transforms into ordinary $ d $-dimensional flat Euclidean space, for which there is a well-known  universal spectral representation for the vacuum mean value of the product of two energy-momentum tensors \cite{Cappelli:1990yc} (see also \cite{Smolkin:2014hba, prep1})
\begin{eqnarray}
&&\langle \hat{T}_{\alpha\beta} (x) \hat{T}_{\rho\sigma} (x') \rangle = \nonumber\\ &&=\frac{A_d}{(d-1)^2} \int_0^{\infty} d\mu\, c^{(0)}(\mu) \Pi^{(0)}_{\alpha\beta,\rho\sigma}(\partial) G_d (x-x',\mu) \nonumber\\
&&+\frac{A_d}{(d-1)^2} \int_0^{\infty} d\mu\, c^{(2)}(\mu) \Pi^{(2)}_{\alpha\beta,\rho\sigma}(\partial) G_d (x-x',\mu)\,,\qquad 
\label{spectral}
\end{eqnarray}
according to which all the differences between the specific theories are accumulated in the spectral densities $ c^{(0)}(\mu) $ and $ c^{(2)}(\mu) $. Here $ A_d= \frac{\Omega_{d-1}}{(d+1)2^{d-1}} $, $ \Omega_{d-1}=\frac{2 \pi^{d/2}}{\Gamma(d/2)} $, $ \Pi^{(0)} $ and $ \Pi^{(2)} $ are the corresponding differential operators of spin-0 and spin-2 parts, and $ G_d (x-x',\mu) $ is the propagator of a scalar field with mass $ \mu $.

In \cite{Smolkin:2014hba, Rosenhaus:2014woa}, the response of the $ \langle \hat{T}_{\mu\nu} \rangle $ to a small deviation of $ \nu $ in (\ref{nu}) from unity was found using the method of expansion in powers of the modular Hamiltonian. In the developed perturbative technique, everything is expressed through the correlators at $ \nu=1 $, which allows to use of the representation (\ref{spectral}). Using the relationship of $ \nu $ with temperature (\ref{nu}), we can immediately obtain from the resulting formulas (6.9) in \cite{Smolkin:2014hba} the derivatives of the components of (\ref{Tgen})
\begin{eqnarray}\nonumber
\frac{\partial \varepsilon}{\partial T}\Big|_a &=& \frac{2 \pi \rho A_d}{(d-1)^2 \Gamma(d)}\int_0^{\infty} d\mu \, \left[\mu^2 K_0(\mu \rho)+\frac{\mu}{\rho} K_1(\mu \rho)\right]\\ \nonumber
&&\cdot \, \left\{c^{(0)}(\mu)+(d-1)(d-2)c^{(2)}(\mu)\right\} \,, \\ \nonumber
\frac{\partial p_\perp}{\partial T}\Big|_a &=& \frac{-2 \pi \rho A_d}{(d-1)^2 \Gamma(d)}\int_0^{\infty} d\mu\, \mu^2 K_0(\mu \rho) \\ \nonumber
&&\cdot \, \left\{c^{(0)}(\mu)-(d-1)c^{(2)}(\mu)\right\} \,, \\ \nonumber
\frac{\partial p_{||}}{\partial T}\Big|_a &=& \frac{2 \pi  A_d}{(d-1)^2 \Gamma(d)}\int_0^{\infty} d\mu\, \mu K_1(\mu \rho)\\
&&\cdot \, \left\{c^{(0)}(\mu)+(d-1)(d-2)c^{(2)}(\mu)\right\}  \,,
\label{solodMain}
\end{eqnarray}
where all derivatives are taken at the point\footnote{All subsequent formulas in the current paper also refer to this case, that is, the Minkowski vacuum.} $ T=T_U $, i.e. $ \nu=1 $, and we use the equality $ \rho=1/a $. For example, one can verify (\ref{solodMain}) using the known spectral functions $ c^{(0)}(\mu) $ and $ c^{(2)}(\mu) $ for massless Dirac field \cite{Smolkin:2014hba}, and comparing with a calculation based on the direct application of (\ref{defder}) to (\ref{12}).

The key observation is the anisotropy following from the equations (\ref{solodMain}). Indeed, the temperature derivatives of the longitudinal and transverse pressures are, generally speaking, not equal for arbitrary spectral densities. Requiring (that is, by introducing that as a sort of  criterion) isotropy
\begin{eqnarray}
\frac{\partial p_{||}}{\partial T} = \frac{\partial p_{\perp}}{\partial T}\,,
\label{sum_phys}
\end{eqnarray}
we obtain from (\ref{solodMain}) the equation
\begin{eqnarray}
\int_0^{\infty} d\mu \left\{ c^{(0)}(\mu) \mathcal{A}^{(0)}(\mu,\rho) + c^{(2)}(\mu) \mathcal{A}^{(2)}(\mu,\rho)\right\}=0\,,\quad
\label{sum}
\end{eqnarray}
with
\begin{eqnarray}\nonumber
\mathcal{A}^{(2)}(\mu,\rho) &=& -(d-1)\mu^2 K_0(\mu \rho)+(d-1)(d-2)\frac{\mu}{\rho} K_1(\mu \rho)\,, \\
\mathcal{A}^{(0)}(\mu,\rho) &=& \mu^2 K_0(\mu \rho)+\frac{\mu}{\rho} K_1(\mu \rho)\,, \nonumber \\
f(\mu,\rho) &=& c^{(0)}(\mu) \mathcal{A}^{(0)}(\mu,\rho) + c^{(2)}(\mu) \mathcal{A}^{(2)}(\mu,\rho)\,.
\label{}
\end{eqnarray}

Condition (\ref{sum}) identifies a class of theories for which radiation in the Rindler space is isotropic. It relates the integrals of the spectral densities $ c^{(0)}(\mu) $ and $ c^{(2)}(\mu) $ and can be considered as a novel sum rule for spectral functions. As we show below, this sum rule holds for any conformal field theory and for massive Dirac fields in any number of dimensions. At the same time, it is nothing but a reformulation of the familiar Pascal's law and can be considered as (one of) the signatures of the hydrodynamic behavior. This isotropy sum rule can also be compared, in particular, with the Burkhardt-Cottingham sum rule \cite{Burkhardt:1970ti} playing the important role in non-perturbative QCD \cite{Efremov:1981sh,Jaffe:1989xx, Soffer:1992ck, Soffer:1997eq}, which is also closely related to rotational symmetry. The latter property was 
stressed in the famous textbook \cite{Feynman:1973xc} by R.P. Feynman and is clearly manifested in QCD factorization \cite{Efremov:1981sh} when the integration over momentum fraction $x$ makes the 
operator local, so that the difference between longitudinal and transverse directions disappears. 

Below we will demonstrate, that by taking into account the sum rule (\ref{sum}), the ratios $\eta/s$ and $\zeta/\eta$ can be obtained in a surprisingly simple way.

\section{Local ratios with viscosities}
\label{sec_etas_new}

\subsection{Shear viscosity}

In \cite{prep1}, the shear viscosity of thermal radiation in the Rindler space was found, using the spectral representation (\ref{spectral}). It is expressed through the spectral function $ c^{(2)}(\mu) $ and depends on the distance from the horizon $ \rho $
\begin{eqnarray}
\eta (\rho) = k_d \rho \int_0^{\infty}d\mu\, c^{(2)}(\mu)\mu^2 K_0(\mu\rho)\,,
\label{etamain}
\end{eqnarray}
where $ k_d=\frac{A_d}{2 \Gamma(d)}=\frac{\pi^{d/2}}{(d+1)2^{d-1}\Gamma(d)\Gamma(d/2)} $. The non-negativity of the spectral function $ c^{(2)}(\mu) \geq 0 $, fixed by unitarity (in Minkowski space), ensures thermodynamic irreversibility in this case through the non-negativity of $ \eta \geq 0 $, demonstrating a close relationship with the renormalization group irreversibility \cite{Cappelli:1990yc, Zamolodchikov:1986gt}. A similar situation holds also for bulk viscosity. as it will be shown below, 

The specific heat can be found from (\ref{solodMain}) and (\ref{defder}). When we  additionally impose the isotropy sum rule on the system (\ref{sum}), the specific heat is expressed only through the spectral function $ c^{(2)}(\mu)$
\begin{eqnarray}
c_V (\rho) = \frac{2 \pi \rho A_d  }{\Gamma(d)}\int_0^{\infty} d\mu\, c^{(2)}(\mu) \mu^2 K_0(\mu \rho)\,. 
\label{cviso}
\end{eqnarray}
Comparing (\ref{etamain}) and (\ref{cviso}) we obtain (\ref{main1})
\begin{eqnarray}
\frac{\eta}{c_V}=\frac{1}{4 \pi}\,.
\label{main_eta_cv}
\end{eqnarray}
Equation (\ref{main_eta_cv}) is actually the most general one. It is useful to rewrite it in terms of entropy density and speed of sound. 
Under the sum rule (\ref{sum_phys}), the derivatives of the different pressure components coincide. This allows us to find the entropy density from (\ref{defder}) and, for example, the second relation in (\ref{solodMain})
\begin{eqnarray} \nonumber
s(\rho) =
 \frac{2 \pi \rho A_d}{(d-1)^2 \Gamma(d)}\int_0^{\infty} d\mu\, \mu^2 K_0(\mu \rho) \\
\cdot \, \left\{(d-1)c^{(2)}(\mu)-c^{(0)}(\mu)\right\} \,.
\label{smain}
\end{eqnarray}
Note that (\ref{smain}) is in accordance with a slightly different approach based on the conformal anomaly \cite{prep1}, as well as with the special cases \cite{Lapygin:2025zhn}.

Substituting (\ref{defder}) into (\ref{main_eta_cv}), we obtain (\ref{main})
\begin{eqnarray}
\frac{\eta}{s}= \frac{1}{4 \pi c_s^2}\,.
\label{eta_s_cs}
\end{eqnarray}

It is interesting to note that the obtained result can be compared to the spirit of holographic calculations, where central charges play important role \cite{Kovtun:2008kx, Buchel:2008vz, Sinha:2009ev}. Remarkably, the example with entanglement clearly shows the source of the relationship between viscosity and entropy (or specific heat): both quantities turn out to be of the same nature, being a response to a change in geometry - in one case to off-diagonal fluctuations of the metric, in the other to a small angular deficit - in both cases governed by $\langle \hat{T}\hat{T} \rangle$ correlators.

%

\subsection{Bulk viscosity}

Entanglement bulk viscosity was also derived in \cite{prep1} from (\ref{spectral}) and is expressed in terms of $ c^{(0)}(\mu) $
\begin{eqnarray}
\zeta (\rho) = \frac{2k_d \rho}{(d-1)^2}  \int_0^{\infty}d\mu\, c^{(0)}(\mu)\mu^2 K_0(\mu\rho)\,.
\label{zetamain}
\end{eqnarray}
Next, we again assume the validity of the sum rule (\ref{sum_phys}), (\ref{sum}), which allows us to determine the speed of sound according to (\ref{defder}). Comparing (\ref{zetamain}), (\ref{etamain}), (\ref{defder}), (\ref{cviso}) and (\ref{smain}), we immediately obtain (\ref{zeta_eta_intro})
\begin{eqnarray}
\frac{\zeta}{\eta} = 2\left(\frac{1}{d-1}-c_s^2\right)\,.
\label{zeta_eta}
\end{eqnarray}

\section{Discussion and applications}
\label{sec_disc}
\subsection{Causality and KSS bound}
\label{secsub_kss}

Comparing (\ref{eta_s_cs}) and (\ref{kss}), we can conclude that KSS bound may be considered as following from causality. Indeed, since the speed of sound is less than the speed of light $ c_s^2<1 $, then $ \eta/s>1/(4\pi) $ from (\ref{eta_s_cs}). It's interesting to compare this with the result \cite{Brigante:2008gz}, which considered viscosity for conformal field theories with Gauss-Bonnet gravity dual. It was shown that $\eta/s < \frac{16}{25} \cdot \frac{1}{4\pi}$ occurs only when superluminal gravitons appear in the dual theory. Our result is similar, although we work outside the holographic context, and  instead of gravitons, the violation could be caused by superluminal phonons. Finally, it should be emphasized that, unlike \cite{Brigante:2008gz}, we do not have the factor $ 16/25 $.

%

\subsection{Tests of the isotropy sum rule}
\label{secsub_valid_sum}
%

\subsubsection{Conformal field theory}
\label{secsub_cft}

For conformal field theory we have \cite{Cappelli:1990yc, Smolkin:2014hba}
\begin{eqnarray}
c^{(0)}(\mu)\sim \mu^{d-2} \delta(\mu)\,,\quad c^{(2)}(\mu)=  \frac{(d-1)}{d}  C_T \mu^{d-3} \,,
\label{conf}
\end{eqnarray}
where $ C_T $ is the conformal central charge.
Substituting (\ref{conf}) into (\ref{sum}), it is easy to verify that (\ref{sum}) holds for any $ d>2 $, since the integrand reduces to the total derivative.
Figure \ref{fig} shows the behavior of the integrand in (\ref{sum}) in this case.

\subsubsection{Massive free Dirac field}
\label{secsub_dirac}

\begin{figure}[!h]
\begin{minipage}{0.5\textwidth}
  \centerline{\includegraphics[width=1\textwidth]{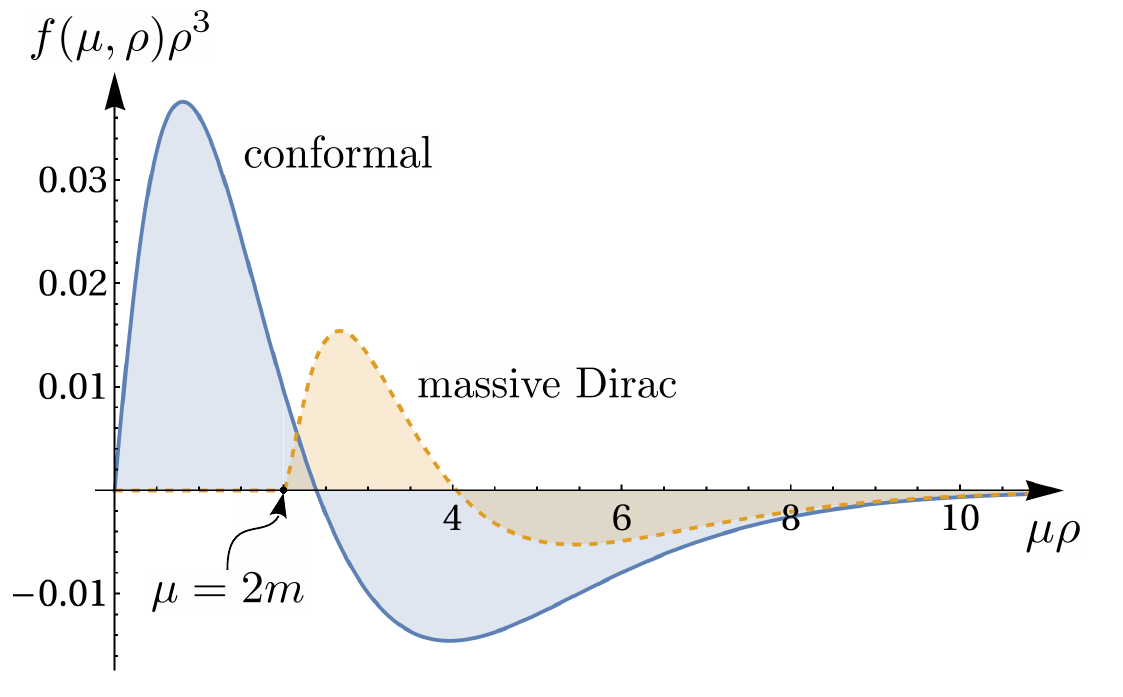}}
\end{minipage}
\caption{Integrand of the sum rule (\ref{sum}) for $d=4$. 
Solid (blue): conformal field theory; 
dashed (orange): free massive Dirac field (as an example, the mass is set equal to $m=1/\rho$).
The vanishing integral ensures the isotropy sum rule. 
Quantities are made dimensionless using $\rho$.}
\label{fig}
\end{figure}

The spectral functions for free massive Dirac fields are \cite{Cappelli:1990yc}\footnote{Compared to  \cite{Cappelli:1990yc}, an additional normalization factor $ \Gamma(d/2)^2/(4 \pi^d) $ is introduced in accordance with the normalization of the central charge in \cite{Erdmenger:1996yc} and \cite{Smolkin:2014hba}.}
\begin{eqnarray}\nonumber
c^{(0)}(\mu, m) &=& \frac{2^{\lfloor d/2\rfloor}(d^2-1) \Gamma\left(\frac{d}{2}\right)^{2}}{2 \pi^d} m^2 \mu^{d-5}   \biggl(1 - \frac{4m^2}{\mu^2}\biggr)^{\!\frac{d-1}{2}}
 \\ \nonumber && \cdot \,\theta (\mu-2 m)\,, \\ \nonumber
c^{(2)}(\mu, m) &=& \frac{2^{\lfloor d/2\rfloor}(d-1) \Gamma\left(\frac{d}{2}\right)^{2}}{8 \pi^d} \mu^{d-3}   \biggl(1 - \frac{4m^2}{\mu^2}\biggr)^{\!\frac{d-1}{2}} \\ 
&&\cdot\left(1 + \frac{8}{d-1} \cdot \frac{m^2}{\mu^2}\right) \theta (\mu-2 m)\,.
\label{cdir}
\end{eqnarray}
Substituting (\ref{cdir}) into (\ref{sum}), we can verify that (\ref{sum}) is satisfied, since the integrand again reduces to a derivative, and the integral equals zero for any $ d>2 $. The graph of the integrand is shown in Figure \ref{fig}.

The fulfillment of the sum rule for the massive Dirac field corresponds to the known isotropy of radiation in this case \cite{Bezerra:2006nu, Prokhorov:2019sss, Buzzegoli:2017cqy}.

\subsubsection{Massive free scalar field}
\label{secsub_scalar}

The non-triviality of condition (\ref{sum}) is emphasized by the existence of a counterexample. The spectral functions for massive scalar field are given in \cite{Cappelli:1990yc} in Eq. (3.32) and it is easy to verify that (\ref{sum}) is violated for them. This is not accidental -- it is known that for a massive scalar field the energy-momentum tensor contains a term $ T_{\mu\nu} \sim a_{\mu} a_{\nu} $ along the acceleration four-vector, that makes an additional contribution to the longitudinal pressure  \cite{Buzzegoli:2017cqy, Iellici:1997ud}. The origins of this anisotropy appear to be an interesting mystery for future research.

\subsection{Possible phenomenological applications}
\label{secsub_phenom}

While the equations (\ref{main}) and (\ref{main1}) were derived for a particular case (accelerated system at $ T=T_U $), they are expressed in terms of general hydrodynamic quantities. One may therefore
speculate that they might represent a lower bound, potentially relevant beyond this specific context \footnote{Though in our case Eqs. (\ref{main}) and (\ref{main1}) are equivalent, (\ref{main1}) is more general, since (\ref{main}) was obtained from (\ref{main1}) using definitions (\ref{defder}) (which are not entirely universal).}.

As is well known, quark-gluon plasma is a fluid that is closest to the KSS-bound (\ref{kss}). 
In this case, the ratio $ \eta/s $ has been studied in detail in numerous studies, based on experimental data, and it is approximately equal to $0.05 < \eta/s < 0.2$ \cite{Harris:2023tti}. The corresponding uncertainties are quite large, and it's possible both that $ \eta/s $ practically saturates the bound $ 1/(4\pi) \sim 0.08 $ or exceeds it several times. From this perspective, it's interesting (very naively) to compare these estimates with (\ref{main}). Taking $ c_s^2\sim 1/3 $ as a reasonable estimate, we obtain $ \eta/s \sim 0.24 $.

At the same time, the entanglement viscosities (\ref{etamain}) and (\ref{zetamain}), describing transport in an accelerated medium, may also be of independent interest. Indeed, transport phenomena and related effects in accelerated and/or rotating media are currently being actively studied in the context of physics in heavy ion collisions \cite{STAR:2017ckg, Son:2009tf, Landsteiner:2011cp, Prokhorov:2022udo, Rogachevsky:2010ys, Kharzeev:2015znc, Tsegelnik:2026ulr, Braguta:2026nfy}. 
Moreover, given that the viscosities (\ref{etamain}), (\ref{zetamain}) are roughly estimated to be $ \eta,\zeta \sim a^3 $ in four dimensions, and since the acceleration 
$ a $ is predicted to be extremely high in heavy-ion collisions \cite{Prokhorov:2025vak, Karpenko:2018erl, Chernodub:2024wis}, one could expect their non-negligible contributions. which reuires the further investigations.

\section{Conclusions}
\label{sec_concl}

We have shown that the general spectral representation results in the ratios of the shear viscosity of thermal radiation in Rindler space to the entropy density and to the specific heat locally satisfy expressions (\ref{main}) and (\ref{main1}). The resulting expression demonstrates the connection between the KSS bound and causality. Furthermore, we demonstrated that the bulk viscosity of the this radiation saturates another famous bound for bulk viscosity.

By imposing the condition of isotropy of thermal radiation in the Rindler space, we find a novel sum rule relating the spin-0 and spin-2 spectral functions, and explicitly verify it for conformal field theories and free massive Dirac fields in any number of dimensions. Finally, we discussed possible phenomenological consequences.


{\bf Acknowledgements}

The authors are thankful to Francesco Becattini, Anastasia A. Golubtsova, Alexei B. Larionov, Dmitri N. Voskresensky and Valentin I. Zakharov for stimulating discussions and comments. The work was supported by Russian Science Foundation Grant No. 25-22-00887.

\bibliography{lit}

\begin{thebibliography}{10}

\bibitem{Policastro:2001yc}
G.~Policastro, Dan~T. Son, and Andrei~O. Starinets.
\newblock {The Shear viscosity of strongly coupled N=4 supersymmetric
  Yang-Mills plasma}.
\newblock {\em Phys. Rev. Lett.}, 87:081601, 2001.

\bibitem{Kovtun:2004de}
P.~Kovtun, Dan~T. Son, and Andrei~O. Starinets.
\newblock {Viscosity in strongly interacting quantum field theories from black
  hole physics}.
\newblock {\em Phys. Rev. Lett.}, 94:111601, 2005.

\bibitem{Chiofalo:2025jph}
Maril{\`u} Chiofalo, Dario Grasso, Stefano Liberati, and Massimo Mannarelli.
\newblock {Shear viscosity to entropy density ratio: A powerful tool for
  gravity theories and strongly coupled fluids}.
\newblock {\em EPL}, 151(6):69001, 2025.

\bibitem{Brigante:2008gz}
Mauro Brigante, Hong Liu, Robert~C. Myers, Stephen Shenker, and Sho Yaida.
\newblock {The Viscosity Bound and Causality Violation}.
\newblock {\em Phys. Rev. Lett.}, 100:191601, 2008.

\bibitem{Damour:1979wya}
Thibaut Damour.
\newblock {\em {Quelques proprietes mecaniques, electromagnet iques,
  thermodynamiques et quantiques des trous noir}}.
\newblock PhD thesis, Paris U., VI-VII, 1979.

\bibitem{Parikh:1997ma}
Maulik Parikh and Frank Wilczek.
\newblock {An Action for black hole membranes}.
\newblock {\em Phys. Rev. D}, 58:064011, 1998.

\bibitem{Chirco:2010xx}
Goffredo Chirco, Christopher Eling, and Stefano Liberati.
\newblock {The universal viscosity to entropy density ratio from entanglement}.
\newblock {\em Phys. Rev. D}, 82:024010, 2010.

\bibitem{Lapygin:2025zhn}
Dmitry~D. Lapygin, Georgy~Yu. Prokhorov, Oleg~V. Teryaev, and Valentin~I.
  Zakharov.
\newblock {Viscosity, entanglement, and acceleration}.
\newblock {\em Phys. Rev. D}, 112(6):065012, 2025.

\bibitem{Unruh:1976db}
W.~G. Unruh.
\newblock {Notes on black hole evaporation}.
\newblock {\em Phys. Rev.}, D14:870, 1976.

\bibitem{Bisognano:1976za}
J.~J Bisognano and E.~H. Wichmann.
\newblock {On the Duality Condition for Quantum Fields}.
\newblock {\em J. Math. Phys.}, 17:303--321, 1976.

\bibitem{Witten:2024upt}
Edward Witten.
\newblock {Introduction to black hole thermodynamics}.
\newblock {\em Eur. Phys. J. Plus}, 140(5):430, 2025.

\bibitem{Srednicki:1993im}
Mark Srednicki.
\newblock {Entropy and area}.
\newblock {\em Phys. Rev. Lett.}, 71:666--669, 1993.

\bibitem{Bombelli:1986rw}
Luca Bombelli, Rabinder~K. Koul, Joohan Lee, and Rafael~D. Sorkin.
\newblock {A Quantum Source of Entropy for Black Holes}.
\newblock {\em Phys. Rev. D}, 34:373--383, 1986.

\bibitem{prep1}
G.~Yu. Prokhorov.
\newblock {Entanglement Viscosity: from Unitarity to Irreversibility in
  Accelerated Frames}.
\newblock {\em arXiv, 2601.02083}, 1 2026.

\bibitem{Zamolodchikov:1986gt}
A.~B. Zamolodchikov.
\newblock {Irreversibility of the Flux of the Renormalization Group in a 2D
  Field Theory}.
\newblock {\em JETP Lett.}, 43:730--732, 1986.

\bibitem{Cappelli:1990yc}
Andrea Cappelli, Daniel Friedan, and Jose~I. Latorre.
\newblock {C theorem and spectral representation}.
\newblock {\em Nucl. Phys. B}, 352:616--670, 1991.

\bibitem{Ellis:2007ic}
George Ellis, Roy Maartens, and Malcolm A.~H. MacCallum.
\newblock {Causality and the speed of sound}.
\newblock {\em Gen. Rel. Grav.}, 39:1651--1660, 2007.

\bibitem{Kovtun:2019hdm}
Pavel Kovtun.
\newblock {First-order relativistic hydrodynamics is stable}.
\newblock {\em JHEP}, 10:034, 2019.

\bibitem{Buchel:2007mf}
Alex Buchel.
\newblock {Bulk viscosity of gauge theory plasma at strong coupling}.
\newblock {\em Phys. Lett. B}, 663:286--289, 2008.

\bibitem{Dowker:1994fi}
J.~S. Dowker.
\newblock {Remarks on geometric entropy}.
\newblock {\em Class. Quant. Grav.}, 11:L55--L60, 1994.

\bibitem{Solodukhin:2011gn}
Sergey~N. Solodukhin.
\newblock {Entanglement entropy of black holes}.
\newblock {\em Living Rev. Rel.}, 14:8, 2011.

\bibitem{Buzzegoli:2017cqy}
M.~Buzzegoli, E.~Grossi, and F.~Becattini.
\newblock {General equilibrium second-order hydrodynamic coefficients for free
  quantum fields}.
\newblock {\em JHEP}, 10:091, 2017.
\newblock [Erratum: JHEP07,119(2018)].

\bibitem{Becattini:2017ljh}
F.~Becattini.
\newblock {Thermodynamic equilibrium with acceleration and the Unruh effect}.
\newblock {\em Phys. Rev.}, D97(8):085013, 2018.

\bibitem{Frolov:1987dz}
Valeri~P. Frolov and E.~M. Serebryanyi.
\newblock {Vacuum Polarization in the Gravitational Field of a Cosmic String}.
\newblock {\em Phys. Rev.}, D35:3779--3782, 1987.

\bibitem{Prokhorov:2019cik}
George~Y. Prokhorov, Oleg~V. Teryaev, and Valentin~I. Zakharov.
\newblock {Unruh effect for fermions from the Zubarev density operator}.
\newblock {\em Phys. Rev.}, D99(7):071901(R), 2019.

\bibitem{Becattini:2023ouz}
Francesco Becattini, Asaad Daher, and Xin-Li Sheng.
\newblock {Entropy current and entropy production in relativistic spin
  hydrodynamics}.
\newblock {\em Phys. Lett. B}, 850:138533, 2024.

\bibitem{Laine:2016hma}
Mikko Laine and Aleksi Vuorinen.
\newblock {Basics of Thermal Field Theory}.
\newblock {\em Lect. Notes Phys.}, 925:pp.1--281, 2016.

\bibitem{Cleymans:2011fx}
J.~Cleymans and D.~Worku.
\newblock {The Hagedorn temperature Revisited}.
\newblock {\em Mod. Phys. Lett. A}, 26:1197--1209, 2011.

\bibitem{Smolkin:2014hba}
Michael Smolkin and Sergey~N. Solodukhin.
\newblock {Correlation functions on conical defects}.
\newblock {\em Phys. Rev. D}, 91(4):044008, 2015.

\bibitem{Rosenhaus:2014woa}
Vladimir Rosenhaus and Michael Smolkin.
\newblock {Entanglement Entropy: A Perturbative Calculation}.
\newblock {\em JHEP}, 12:179, 2014.

\bibitem{Burkhardt:1970ti}
Hugh Burkhardt and W.~N. Cottingham.
\newblock {Sum rules for forward virtual Compton scattering}.
\newblock {\em Annals Phys.}, 56:453--463, 1970.

\bibitem{Efremov:1981sh}
A.~V. Efremov and O.~V. Teryaev.
\newblock {On Spin Effects in Quantum Chromodynamics}.
\newblock {\em Sov. J. Nucl. Phys.}, 36:140, 1982.

\bibitem{Jaffe:1989xx}
R.~L. Jaffe.
\newblock {$g_{2}$-The Nucleon's Other Spin-Dependent Structure Function}.
\newblock {\em Comments Nucl. Part. Phys.}, 19(5):239--257, 1990.

\bibitem{Soffer:1992ck}
Jacques Soffer and O.~Teryaev.
\newblock {The Role of g-2 in relating the Schwinger and Gerasimov-Drell-Hearn
  sum rules}.
\newblock {\em Phys. Rev. Lett.}, 70:3373--3375, 1993.

\bibitem{Soffer:1997eq}
Jacques Soffer and O.~V. Teryaev.
\newblock {Comment on the Burkhardt-Cottingham and generalized
  Gerasimov-Drell-Hearn sum rules for the neutron}.
\newblock {\em Phys. Rev. D}, 56:7458--7460, 1997.

\bibitem{Feynman:1973xc}
Richard~P. Feynman.
\newblock {\em Photon-{Hadron} {Interactions}}.
\newblock Advanced {Books} {Classics}. Chapman and Hall/CRC, Boulder, 2018.

\bibitem{Kovtun:2008kx}
Pavel Kovtun and Adam Ritz.
\newblock {Universal conductivity and central charges}.
\newblock {\em Phys. Rev. D}, 78:066009, 2008.

\bibitem{Buchel:2008vz}
Alex Buchel, Robert~C. Myers, and Aninda Sinha.
\newblock {Beyond eta/s = 1/4 pi}.
\newblock {\em JHEP}, 03:084, 2009.

\bibitem{Sinha:2009ev}
Aninda Sinha and Robert~C. Myers.
\newblock {The Viscosity bound in string theory}.
\newblock {\em Nucl. Phys. A}, 830:295C--298C, 2009.

\bibitem{Erdmenger:1996yc}
J.~Erdmenger and H.~Osborn.
\newblock {Conserved currents and the energy momentum tensor in conformally
  invariant theories for general dimensions}.
\newblock {\em Nucl. Phys. B}, 483:431--474, 1997.

\bibitem{Bezerra:2006nu}
Valdir~B. Bezerra and Nail~R. Khusnutdinov.
\newblock {Vacuum expectation value of the spinor massive field in the cosmic
  string space-time}.
\newblock {\em Class. Quant. Grav.}, 23:3449--3462, 2006.

\bibitem{Prokhorov:2019sss}
Georgy~Y. Prokhorov, Oleg~V. Teryaev, and Valentin~I. Zakharov.
\newblock {Calculation of acceleration effects using the Zubarev density
  operator}.
\newblock {\em Particles}, 3(1):1--14, 2020.

\bibitem{Iellici:1997ud}
Devis Iellici.
\newblock {Massive scalar field near a cosmic string}.
\newblock {\em Class. Quant. Grav.}, 14:3287--3301, 1997.

\bibitem{Harris:2023tti}
John~W. Harris and Berndt M{\"u}ller.
\newblock {''QGP Signatures'' Revisited}.
\newblock {\em Eur. Phys. J. C}, 84(3):247, 2024.

\bibitem{STAR:2017ckg}
L.~Adamczyk et~al.
\newblock {Global $\Lambda$ hyperon polarization in nuclear collisions:
  evidence for the most vortical fluid}.
\newblock {\em Nature}, 548:62--65, 2017.

\bibitem{Son:2009tf}
Dam~T. Son and Piotr Surowka.
\newblock {Hydrodynamics with Triangle Anomalies}.
\newblock {\em Phys. Rev. Lett.}, 103:191601, 2009.

\bibitem{Landsteiner:2011cp}
Karl Landsteiner, Eugenio Megias, and Francisco Pena-Benitez.
\newblock {Gravitational Anomaly and Transport}.
\newblock {\em Phys. Rev. Lett.}, 107:021601, 2011.

\bibitem{Prokhorov:2022udo}
G.~Yu. Prokhorov, O.~V. Teryaev, and V.~I. Zakharov.
\newblock {Hydrodynamic Manifestations of Gravitational Chiral Anomaly}.
\newblock {\em Phys. Rev. Lett.}, 129(15):151601, 2022.

\bibitem{Rogachevsky:2010ys}
Oleg Rogachevsky, Alexander Sorin, and Oleg Teryaev.
\newblock {Chiral vortaic effect and neutron asymmetries in heavy-ion
  collisions}.
\newblock {\em Phys. Rev.}, C82:054910, 2010.

\bibitem{Kharzeev:2015znc}
D.~E. Kharzeev, J.~Liao, S.~A. Voloshin, and G.~Wang.
\newblock {Chiral magnetic and vortical effects in high-energy nuclear
  collisions - A status report}.
\newblock {\em Prog. Part. Nucl. Phys.}, 88:1--28, 2016.

\bibitem{Tsegelnik:2026ulr}
Nikita Tsegelnik.
\newblock {The Static Heavy Quark-Antiquark Potential within String Theory in
  Arbitrary Stationary Backgrounds}.
\newblock {\em arXiv, 2601.10668}, 1 2026.

\bibitem{Braguta:2026nfy}
Viktor Braguta, Vladimir Goy, Jayanta Dey, and Artem Roenko.
\newblock {Spatial confinement-deconfinement transition in accelerated
  gluodynamics within lattice simulation}.
\newblock {\em arXiv, 2602.20970}, 2 2026.

\bibitem{Prokhorov:2025vak}
G.~Yu. Prokhorov, D.~A. Shohonov, O.~V. Teryaev, N.~S. Tsegelnik, and V.~I.
  Zakharov.
\newblock {Modeling of acceleration in heavy-ion collisions: Occurrence of
  temperature below the Unruh temperature}.
\newblock {\em Phys. Rev. C}, 112(6):064907, 2025.

\bibitem{Karpenko:2018erl}
Iurii Karpenko and Francesco Becattini.
\newblock {Lambda polarization in heavy ion collisions: from RHIC BES to LHC
  energies}.
\newblock {\em Nucl. Phys.}, A982:519--522, 2019.

\bibitem{Chernodub:2024wis}
M.~N. Chernodub, V.~A. Goy, A.~V. Molochkov, D.~V. Stepanov, and A.~S.
  Pochinok.
\newblock {Extreme Softening of QCD Phase Transition under Weak Acceleration:
  First-Principles Monte~Carlo Results for Gluon Plasma}.
\newblock {\em Phys. Rev. Lett.}, 134(11):111904, 2025.

\end{thebibliography}

\end{document}